\documentclass[11pt,english]{article}
\usepackage[T1]{fontenc}
\usepackage[latin1]{inputenc}
\usepackage{geometry}
\usepackage{amsmath}
\usepackage{graphicx}
\usepackage{setspace}
\usepackage{amssymb}
\usepackage[authoryear]{natbib}
\usepackage{rotating}

\makeatletter

\providecommand{\tabularnewline}{\\}

\usepackage{natbib}
\usepackage{fancyhdr}

\usepackage{babel}
\makeatother

\begin{document}
\begin{center}Internal Structure of Massive Terrestrial
Planets\\
\end{center}

\begin{center}Diana Valencia$^{a,1}$, Richard J O'Connell$^{a}$
and Dimitar Sasselov$^{b}$\\
\bigskip

\end{center}

\begin{center}$^{a}$ Earth and Planetary Sciences, Harvard University,
Cambridge, Massachussets, 02138\\
$^{b}$ Harvard-Smithsonian Center for Astrophysics, Department of
Astronomy, Harvard University, Cambridge, Massachussets, 02138\end{center}

\begin{center}$^{1}$Corresponding Author E-mail address: valencia@mail.geophysics.harvard.edu\\
\end{center}

\bigskip
\bigskip

\textbf{ABSTRACT}
\bigskip

Planetary formation models predict the existence of massive terrestrial
planets and experiments are now being designed that should succeed in 
discovering them and measuring their masses and radii. 
We calculate internal structures of planets with one to ten
times the mass of the Earth (Super-Earths) in order to obtain scaling
laws for total radius, mantle thickness, core size and average density
as a function of mass. We explore different compositions and obtain a scaling
law of $R\propto M^{0.267-0.272}$ for Super-Earths. We also study a
second family of planets, Super-Mercuries with masses ranging from
one mercury-mass to ten mercury-masses with similar composition to
the Earth's but larger core mass fraction. We explore the effect of
surface temperature and core mass fraction on the scaling laws for
these planets. The scaling law obtained for the Super-Mercuries is
$R\propto M^{\sim0.3}$.\\


\pagebreak

\section{Introduction}

In the last decade, astronomers have indirectly detected 
more than 160 planets orbiting stars other than our Sun. The two methods 
used most extensively are 1) the radial velocity method that uses
the Doppler shift in the parent star's motion caused by the orbiting
planet and 2) the method of transits that uses the periodic drop in
brightness of the star as the planet eclipses it in its motion
around it. There is very little known about these new planets, since
no direct observations have been made  yet, however in at least
one case the transit observations led to the detection of the planet's
atmosphere (\citealt{Charbonneau:2002}, \citealt{Vidal-Madjar:2003}). 
Also, by combining
the two methods one can determine both the planet's mass and radius,
yielding some insight into their interior structure. For seven 
extrasolar planets in the mass range of $0.5-1.5~M_{Jup}$ the estimated
mean bulk densities range from $0.4-1.3~g~cm^{-3}$ (\citealt{Konacki:2005}).
These imply gas giant planet structure similar to
that of Jupiter and Saturn.\\

Up to this point all extrasolar planets discovered have been gas giants.
A large fraction of them are in very tight orbits (with periods of a few
days) making their surface temperature very hot (Hot Jupiters). Only in
the past year have discoveries reached masses as small as those of Neptune
and Uranus. Certainly all search methods are biased to detect
the largest and most massive planets, but the frontier is being quickly
pushed down using both higher precision as well as a larger and diverse
sample of stars. Space missions under development, such as
Kepler (\citealt{Kepler-Mission}) and COROT (\citealt{COROT:2002}),
are fully capable (and expected) to discover many massive terrestrial
planets (Super-Earths); the Kepler mission expects to discover almost
one thousand Super-Earths.  Such expectations are based in part on
theoretical work that shows no evidence precluding planet formation
in the $1-10~M_{\oplus}$ mass range (e.g. \citealt{Ida:2004}), even
in planetary systems with Hot Jupiters (\citealt{Lunine:AGU}).
Therefore, we believe it is timely to compute models of the internal
structure of Super-Earths and study the possible mass-radius relations,
especially because we have found no such previous work in the literature.\\

Though, far from complete, our understanding of the Earth is the most
comprehensive that we have of any planet and should be the starting point
to study massive terrestrial planets. In this paper we have scaled
Earth to more massive planets as the first step into understanding
how other terrestrial planets might have accreted and evolved. 
We address the likelihood of existence of these planets in section two. In section
three we describe the method we use to obtain the density distribution,
pressure and temperature profiles of ten planets spanning from one
earth-mass to ten earth-masses with same chemical and mineral composition as the
Earth, which from here on will be called Super-Earths. Scaling laws
are derived for the total radius, mantle thickness, core radius and
average mantle density as the total mass of an Earth-like planet increases.\\

It will become clear in section three that an a-priori assumption on
the ratio of $\text{Fe/Mg}$ needs to be made to account
for the amount of mass of the planet in the core. Therefore, in section
three we consider another family of planets with different $\text{Fe/Mg}$
ratio, corresponding to the ratio believed for Mercury, and with masses
ranging from one mercury-mass to ten mercury-masses (Super-Mercuries),
and we investigate the effect of surface temperature on the scaling
laws.\\

\section{On the Formation of Super-Earths}

Given the obvious lack of terrestrial planets of mass between $1-10~M_{\oplus}$
in our own Solar System, it is appropriate to investigate if such planets are
unlikely to form. As we shall see all indications are that they will.
First, in reconstructing the minimum-mass solar nebula (the proto-planetary disk
that gave rise to the Solar System planets), as well as our current
knowledge of what are typical proto-planetary disks (\citealt{Hayashi:1981},
 \citealt{Wetherill:1990}, \citealt{ Beckwith:1996}, \citealt{Wyatt:2003}), we find plenty of 
solid material to accumulate planets larger than the Earth.
Most generally (including orbit migration), what limits the masses of rocky
planets in the inner parts of a disk (few AU [Astronomical Unit$= a_{\oplus}$]
from the star) is the available amount of solid material (\citealt{Ida:2004}):
$$
M\backsimeq 1.2{\eta}_{ice}f\biggl(\frac{a}{1~AU}\biggr)^{1/2}~M_{\oplus}
$$
where $f$ is a scaling parameter for the total disk mass (typical values are
in the 0.1$-$30 range), ${\eta}_{ice}$ is a step function (jumps from 1 to $\sim$4)
describing the increase in density of solid particles as ice condenses outside
the 'snow line' at an orbital distance, $a$, of a few AU. 

Second, as a planet embryo of mass $M$ grows above about an Earth mass, it begins
to accrete gas from the disk. Such accretion could quickly turn into a runaway
process that would lead to the formation of a gas giant planet like Jupiter
(\citealt{Mizuno:1980}, \citealt{Stevenson:1982}, \citealt{Bodenheimer:1986}).
The value of the critical mass, $M_{crit}$, for which such runaway growth can occur would
thus determine whether Super-Earths are likely to exist at all.

Precise determination of $M_{crit}$ would require 3-dimensional radiation hydrodynamics
models. Current simpler numerical models (\citealt{Ikoma:2000}) find a robust
dependence on the rate of accretion, $\dot{M}$, and the opacity, $\kappa$, of the
disk gas:
$$
M_{crit}\backsimeq 10\biggl(\frac{\dot{M}}{10^{-6}M_{\oplus}yr^{-1}}\biggr)^q
\biggl(\frac{\kappa}{1 cm^2 g^{-1}}\biggr)^s M_{\oplus},
$$
where $q$ and $s$ lie between 0.2 and 0.3. One might expect that planets grow so
rapidly from $\sim$10 to more than 100$M_{\oplus}$ that they would rarely be
found with a final mass in the intermediate range ($10-100~M_{\oplus}$),
as shown by the detailed calculations of Ida \& Lin (2004). These same
calculations predict a large number of Super-Earths in the inner parts of the disk.
It is important to note that a certain fraction of the Super-Earths will have
a substantial volatile content (primarily water); while dry terrestrial planets
are most common in the inner disk, the expected fraction of 'ocean planets' is
poorly known (\citealt{Raymond:2004}).

We hence conclude that the common existence of Super-Earths ($1-10~M_{\oplus}$) is
a reasonable expectation of the current theory of planet formation.

\section{Earth and Super-Earths}

\subsection{Method}

In a radially symmetric planet that is chemically and mineralogically
homogeneous the equations describing the density $\rho$ (Adam-Williamson equation), gravity
$g$, mass $m$ and pressure $P$ derivatives are:

\begin{eqnarray}
\frac{d\rho}{dr} & = & -\frac{\rho(r)g(r)}{\phi(r)}\label{adam_wil}\end{eqnarray}

\begin{equation}
\frac{dg}{dr}=4\pi G\rho(r)-\frac{2Gm(r)}{r^{3}}\label{dg/dr}\end{equation}

\begin{equation}
\frac{dm}{dr}=4\pi r^{2}\rho(r)\label{dm/dr}\end{equation}

\begin{equation}
\frac{dP}{dr}=-\rho(r)g(r)\label{dP/dr}\end{equation}

where $\phi(r)=\frac{K_S(r)}{\rho(r)}$ is the seismic parameter that
can be calculated from equations of state, $K_S$ is the adiabatic bulk modulus,
$G$ is the gravitational constant and $r$ is the distance from the
center of the planet.\\

The equation of state (EOS) we choose is the third order Birch-Murnaghan
equation of state with a Debye thermal correction. This EOS results
from the expression of the Helmholtz free energy as a power series
expansion of finite strain and the thermal correction accounts for
lattice vibrations with a cut-off frequency corresponding to the Debye
frequency. This EOS has been shown to be adequate for our purposes
(\citealt{Birch-1952}, \citealt{Poirier:Earths-interior:EOS}) and
its expression is the following\begin{equation}
K_T(\rho,T)=K_T(\rho,300)+\Delta K_{th}(\rho,T)\label{eosBM}\end{equation}

{\footnotesize \[
K_T(\rho,300)=\frac{K_{0,300}}{2}\left\{ \left(7x^{7/3}-5x^{5/3}\right)\left[1+\frac{3}{4}\left(K_{0,300}'-4\right)\left(x^{2/3}-1\right)\right]+\frac{3}{2}\left(x^{9/3}-x^{7/3}\right)\left(K_{0,300}'-4\right)\right\} \]
}{\footnotesize \par}

\[
\Delta K_{th}=3nR\gamma\rho\left(f(T)-f(T_{0})\right)\]

\[
f(T)=\left(1-q-3\gamma\right)\frac{T^{4}}{\theta^{3}}\int_{0}^{\frac{\theta}{T}}\frac{\xi^{3}d\xi}{\exp\xi-1}+3\theta\gamma\frac{1}{\exp\left(\theta/T\right)-1}\]

\[
K_S=K_T \left\{1+\alpha \gamma T \right\}
\]

where $K_{0,300}$ and $K_{0,300}^{'}$ are the isothermal bulk modulus
and its first derivative at zero pressure and $T_{0}=300$K; $K_T$ and $K_S$ are the isothermal and adiabatic bulk modulus 
at any given radius; $x(r)=\frac{\rho(r)}{\rho_{0}}$
is the ratio of density to the uncompressed density; $\alpha$ is the coefficient of thermal expansion; $R$ is the universal
gas constant; $n$ is the number of atoms in the unit cell; $\theta$
is the Debye temperature and $\gamma$ is the Gruneisen parameter
taken to vary with depth as $\theta=\theta_{0}\exp\left(\frac{\gamma_{0}-\gamma}{q}\right)$ and $\gamma=\gamma_{0}\left(\frac{\rho_{0}}{\rho(r)}\right)^{q}$. This EOS does not include a contribution from electronic pressure in metals known to have a 8\% 
contribution to pressure at 350 GPa compared to a 22\% contribution from thermal pressure (\citealt{Stixrude_Wasserman:1997}). 
Our results are not affected by this term and will be discussed in section 3. \\

The system of PDE's described in Eq. \ref{adam_wil} to \ref{dP/dr}
in conjunction with Eq. \ref{eosBM} is integrated numerically from
the surface using a fourth order Runge-Kutta algorithm with initial
conditions of a given surface density, zero surface pressure, total
mass $M$, and corresponding surface gravity $g_{s}=GM/R^{2}$, where
$R$ is the total radius. The algorithm searches for an appropriate
total radius corresponding to a fixed total mass $M$ and composition
included in the values for $\rho_{0},K_{0,300}$ , $K_{0,300}^{'}$,
$\gamma_{0}$, $q$ and $\theta_{0}$ in the EOS and stops when it
finds a total radius that yields no excess mass in the last inner
shell of the integration. \\

Because the Earth is not homogeneous in composition we need to include mixtures
between different mineral phases in the different regions of the Earth.  There are two approaches 
that can be implemented: to use the thermodynamic values at some reference state (T=300 and atmospheric pressure)
as an equivalent of the mix of the mineral components for each region (upper mantle, transition zone, lower mantle, outer core, inner core),
or mix the different mineral phases at each radius of the integration by performing a weighted by mol linear interpolation of the thermodynamic parameters
(density, bulk modulus and Gruneisen parameter).\\

Since the EOS data applies to a particular material, the latter approach is  more exact and we have used it in the mantle.  
Due to the lack of extensive and complete thermodynamic data
for iron and iron with lighter elements at relevant pressures and temperatures, we use the first approach for the core.
The mineral phases that are considered in the mantle are olivine and high pressure forms of olivine.  For the core we used Fe with 8\% by weight Si, extreme end members FeO and Fe, and
Fe plus some alloy equivalent that best fits the density and bulk modulus data from the PREM model (\citealt{PREM}).
\\

In order to include the phase changes that are known to exist in the
Earth and are expected to be present in the Super-Earths the temperature
profile needs to be calculated. We use simple boundary layer convection
models used to describe mantle convection on Earth (\citealt{Turcotte_Schubert:fluid_mechanics}) to obtain the regions where conduction is present (surface
and core-mantle boundary layers) and where convection is present (bulk
mantle and core). We assume that the surface boundary layer is mobile, as is the case for Earth.
A stagnant lid treatment would yield differences in the sizes of the boundary layers. These differences
have an effect on the exact structure of the planet but not on the exponent of the scaling laws we derive in this study. \\

The temperature gradient through the boundary
layers at the surface and core-mantle boundary (CMB) is 
\begin{equation}
\frac{dT}{dr}=-\frac{q_{planet}}{k}\label{eq:fouriers law}
\end{equation}

where $q_{planet}$ is the heat flux at the surface ($q_{s}$) or
at CMB ($q_{cmb}$) and $k$ is the thermal conductivity. The mantle
is known to be convecting and in this study is taken to be convecting
as a single layer. The core is also known to be in a very vigorous convecting
regime in order to sustain the Earth's dynamo (\citealt{Fearn:1998}). The
convective temperature profiles in the mantle and the core are described by the
adiabatic gradient

\begin{equation}
\frac{dT(r)}{dr}=\frac{\rho(r)g(r)T(r)}{K_S(r)}\gamma(r)\label{eq:adiabatic gradient}\end{equation}

which holds in the convecting region.  At the upper and lower boundaries, heat is transported by conduction
through a boundary layer of thickness $\delta$. The Rayleigh number governing convection
may be expressed in terms of the surface heat flux $q_{s}$ 
\begin{equation}
Ra=\frac{\rho g\alpha q_{s}/k}{\kappa\eta}D^{4}\label{eq:Ra}
\end{equation}

 where $\rho$ is the average upper mantle density, $\alpha$ is the thermal expansivity, $\kappa$ is the thermal
diffusivity and $\eta$ is the viscosity. Considering that the dimensionless heat transport
should be a function of the Rayleigh number and that the thickness
of the boundary layer should not depend on the thickness of the mantle
for a vigorously convecting incompressible system (\citealt{O'Connell-Hager-1980}),
the expression for the thickness of the surface boundary layer is

\begin{equation}
\delta=a\frac{D}{2}\left(\frac{Ra}{Rac}\right)^{-1/4}\label{delta}
\end{equation}

where $a$ is a coefficient of order unity. Boundary layer models give the same result (\citealt{Turcotte_Schubert:fluid_mechanics}).
 We have included the dependency of viscosity on temperature with the relationship
$\eta(T)=\eta_0 (\frac {T}{T_0})^{-n}$ with $n=30$ proposed by (\citealt{Davies:1980}) to be adequate for the Earth. Parameters in Eq.8  are those for the upper mantle consistent with a boundary layer model in which convection is governed by the stability of the boundary layer.\\
 
The heat
driving convection in the Earth is a combination of heat produced
from within by radioactive sources and from cooling of the Earth, including heat conducted from the core
into the mantle at the CMB. The exact contributions are not well known,
thus the thickness of the boundary layer at CMB and the heat conducted down the adiabat in the core are uncertain (\citealt{Labrosse_et_al:2001}). 
 In this study the value chosen ad-hoc
to describe the conductive region in the mantle at CMB is half the
thickness of the surface boundary layer. 
For planets that reach very high pressures, compressibility effects will become important, and hence, this simple
parametric approach will fall short from a true description. In addition, this model assumes that the rheology is relatively uniform,
which may not be the case at very high pressures.  Nevertheless, it is the first and most simple analytic step
to take.
Figure 1 shows a schematic
diagram of the temperature profile in the Earth's mantle.\\
{[}Figure 1{]}\\

The first stage of the numerical modeling includes a planet with
only a core and a mantle (ie. no phase transitions). The core mass
fraction (CMF) of the Earth is specified a priori as 32.59\% (\citealt{AppendixA-Stacey:Physics-Earth})
reflecting the $\text{Fe/Mg }$ratio. The EOS in this stage has no
thermal corrections since no temperature profile has been calculated
yet. \\

The second stage of the modeling calculates the temperature profile
using Eq. \ref{eq:fouriers law}-\ref{delta} with average density
and gravity of the upper layer of the mantle obtained for the planet in the
first stage. The temperature profile is used to determine the locations of all major
known phase transitions in the Earth, from olivine ($\text{Mg}_{1-x},\text{Fe}_{x})_{2}\text{SiO}_{4}$
to a more highly coordinated structure (wadsleyite) that occurs at
$\sim410$ km depth in the Earth; the transition from a high pressure
olivine phase (ringwoodite) to perovskite ($\text{Mg}_{1-x},\text{Fe}_{x})\text{SiO}_{3}$
plus ferromagnesiowustite ($\text{Mg}_{1-x},\text{Fe}_{x})\text{O}$
that occurs at $\sim660$ km in the Earth and defines the lower-upper
mantle boundary; and the inner-outer core phase transition that corresponds
to the depth at which the geotherm intersects the solidus of iron
plus an alloy of lighter elements. The solidus for the iron alloy
is defined using Lindeman's law (\citealt{Poirier:Earths-interior:Melting})
\[
T_{m}=T_{m_{0}}\left(\frac{\rho_{0}}{\rho}\right)^{2/3}\exp\left\{ \frac{2\gamma_{0}}{q}\left[1-\left(\frac{\rho_{0}}{\rho}\right)^{q}\right]\right\} \]

where $T_{m_{0}}$ is the melting temperature at zero pressure. \\

The thickness of the boundary layers and hence, the temperature profile and phase boundary locations,
are calculated iteratively with the gravity and density average values
of the upper mantle until convergence is reached and the planet has
reached a self-consistent temperature and density profile. This procedure
is followed for the ten Super-Earths.\\

\subsection{Results}

Thermodynamic data varies little for mantle materials but varies widely for core materials.  
Additionally, there is more uncertainty in the composition of the inner and outer cores.
We explore different compositional EOS parameters ($\rho_{0},$$K_{0,300},K_{0,300}',\gamma_{0},q,\theta$)
to test the robustness of the scaling results.  
Thermodynamic values for the mantle are taken from the compiled and completed data set of \citet{Stixrude_Lithgow:2005}.  
Data for the core is less accurate and incomplete. We have chosen to use two extreme end-member compositions to model Earth and
Super-Earths of FeO and pure Fe that yield a deficit and an excess in density with respect to PREM of 200 kg/m$^3$ and 300 kg/m$^3$
respetively. In addition we also use a composition of Fe and 8\% by mol Si for the core.  All the different cases and thermodynamic values 
are shown in Table 1. \\

{[}Table 1{]}

In order to address the effects of Fe in the mantle and content of ferromagnesiowustite in the lower mantle we fix the composition of the
core to a mixture of Fe and a lighter alloy in the outer and inner core to fit the PREM model more accurately.  We compute 
mixtures of magnesium and iron end members at  0\%, 10\% and 20\% of Fe by mol the mantle; as well as the effect of $(\text{Mg}_{1-x}, \text{Fe}_{x}) O$ 
by including 10\% 30\% and 50\% by mol in the lower mantle.  $\gamma_{0}, T_{m_{0}}\text{ and }q$  for the inner core alloy
are chosen within reasonable values, such that the melting transition for a one
earth-mass planet happens within $30$ kms of the inner-outer core boundary
in the Earth as described by the PREM model.\\

There is a wide range of values for the first and second Gruneisen parameters.  Large first Gruneisen values yield 
a hotter interior with steeper solidus.  The second Gruneisen parameter controls the curvature of the melting regime. 
Certain combinations of these parameters, especially with large second Gruneisen values,
with a given pressure profile lead to a negative P-T slope in the solidus that is not seen in the pressure regime for the Earth.
 This was avoided by carefully choosing the values within the uncertanties in the literature, 
such that the intersection of the thermal profile with the solidus yielded a liquid outer core and a solid inner core. 
 The freedom in these parameters does not affect the values of the scaling laws, mainly because of the smaller effect of temperature. 
Nevertheless, they do affect structure details such as the exact size of the solid inner core or the 
presence of a liquid core for planets more massive than the Earth.\\ 

These cases exhibit a family of planets with complete solid cores except for Earth. The geotherms we obtained in these cases are
closer to the cold geotherm of \citet{Brown_Shankland:1981} and are a consequence of the
thermodynamic data, in particular the relatively low first Gruneisen parameter and high second Gruneisen parameter (see Fig. 4). The intersection with the solidus happens at a low temperature of $\sim 3700$K, a
consequence of a
cold geotherm. We test the effects of a warmer geotherm by setting the Gruneisen parameters to large values in the mantle and core (see Table 1 last two rows)
to obtain a temperature profile similar to the one proposed by \citet{AppendixG-Stacey:Physics-Earth} for the Earth with an inner core boundary temperature of $\sim$ 5000K.
This yielded all Super-Earths with partially liquid cores.  Our last test is to increase the curvature of the solidus (high second Gruneisen parameter)
to yield a family of Super-Earths with liquid cores.\\

{[}Figure 2{]}\\

Figure 2 shows the density and incompressibility profiles of
PREM in red and the profiles obtained for the different sets of mineralogical
values for one earth-mass. Even though the inner core values for the
incompressibility are larger than the ones inferred in PREM, this discrepancy has little effect on the scaling laws. This
is clear from the results of the family of planets with no solid inner core that have similar scaling laws.\\

For planets with masses larger than the Earth, the composition (including heat sources), surface
temperature and core mass fraction chosen are equal to those for the one earth-mass
case. The heat flow is assumed to scale linearly with mass, as a first attempt
to scale internal heat production and secular cooling of more massive
planets.\begin{equation}
4\pi R^{2}q_{planet}=Q_{planet}=Q_{earth}\frac{M_{planet}}{M_{earth}}\label{Q scale M}\end{equation}
\\

For one composition (10\% Fe in mantle minerals and 30\% $(\text{Mg}_{1-x},\text{Fe}_{x})$O in the
lower mantle), Fig. 3 shows the results for the density profile of
the ten Super-Earths. We also include a power law fit of total radius
with mass for all the different compositions.  Figure 4 shows the corresponding
temperature profile, as well as the iron plus a light alloy melting curve for
these ten planets. It is clear that the only planet that is hot enough
to exhibit an outer liquid core is the Earth. In larger planets the effect of pressure overwhelms
the temperature, to yield a solid metal core. Owing to the strong dependence of viscosity in temperature, the internal
temperature beneath the top boundary layer is almost independent of mass.  Additionally, the thickness of the boundary layer 
decreases with increasing mass as a consequence of higher Rayleigh numbers for more massive planets\\
{[}Figure 3, Figure 4{]}\\

We fit total radius, size of the mantle, core radius and average density
with mass in a power law relationship $\propto M^{\beta}$ for each
of the ten mineral compositions. Figure 5 shows the results of the
scaling coefficient $\beta$ for all the sets, Table 1 includes the results as well as the thermodynamic 
values used in each case. All fits are quite
linear in log space except for the average density relationship that
has a slight curvature. \\

A one earth-mass planet with a core composition of FeO yields a density deficit compared to the Earth, as expected (see Fig.2).  
In this family of planets, the first three exhibit completely liquid cores and the rest have completely 
solid cores. Figure 5 shows the results from this composition in a grey down-pointed triangle.  Due to the lighter
composition in the core, given a fixed core mass fraction, 
more massive planets accomodate more of their mass in their mantle, yielding
a larger scaling exponent for the mantle thickness.  Due to the $r^3$ geometric factor this translates into smaller 
core radius and total radius exponents.  In turn, a smaller exponent for the
total radius translates into a larger exponent for the average density.
Nevertheless, this composition gives extreme values for the scaling of the bulk properties for Earth-like planets
and hence may be regarded as an extreme case.
An FeSi8\% composition yields liquid cores for all the Super-Earths planets. A one earth-mass planet is still too dense to match PREM. 
A core composition of pure Fe yields a one earth-mass planet with an excess in density with respect to PREM
and solid cores for the family of planets.\\

Circles, stars and plus signs in Fig. 5 correspond to
10\%, 30\% and 50\% ferromagnesiowustite in the lower mantle. Wustite is lighter and more compressible than perovskite. 
Hence more wustite in the lower mantle decreases the total radius and mantle thickness scaling exponent.  
However, more wustite changes the scaling of the core radius very little. 
Consequently the scaling for the average density increases with the content of ferromagnesiowustite 
in the lower mantle. 
 \\
{[}Figure 5{]}\\

The values labeled 'hot' in Fig. 5 are the sets in which we chose
high thermal parameters.  For the geotherm that matches that of \citealt{Stacey:Physics-Earth}, the whole family of planets
exhibit a liquid outer core.
The values shown in green and pink asterisks have the same composition of 10\%Fe and 30\% ($\text{Mg}_{1-x},\text{Fe}_x$)O and different geotherms.
The scaling for radius of a warmer interior increases with respect to the cold geotherm case, whereas the core radius scaling 
decreases slightly, producing a larger $\beta$ for mantle thickness
and lower average density. The results shown with a purple asterisk are produced with the same composition and
a high second Gruneisen parameter that allows for all Super-Earths
to have a completely liquid core except for Earth. 
 \\

The scaling parameters do not vary much with these hotter geotherms because of the smaller effect of temperature 
on the density of the planet compared to the pressure effect. Choosing different values for the thickness of the bottom
boundary layer would make the core hotter or colder by no more than $\sim$300K, not enough to yield different scaling laws for the bulk properties of these planets.\\

The effects included to calculate the pressure are the hydrostatic pressure
derived from the 3rd order Birch-Murnaghan equation of state and the thermal pressure derived
from the Debye theory. It is known that the contribution from electronic pressure is non-negligible in metals at high pressures but is still smaller
than the thermal pressure contribution. \citet{Stixrude_Wasserman:1997} showed that at a pressure of 350 GPa, close to the Earth's central pressure, the
electronic pressure has an 8\% contribution.  This is the error in pressure that can be expected for Super-Earths.
Despite this uncertainty, the scaling can be considered to be robust by looking at the results in Table 1 for the cases of
pure FeO and pure Fe in the outer core with a deficit and excess in density (hence in pressure), with respect to PREM, of $\sim 10\%$ and $16\%$ respectively.
Including this term in the pressure would make the core material more incompressible in very massive planets, making it more difficult to accomodate mass in the core.  
Thus, the effect would be to yield comparatively smaller radii with large masses, decreasing the scaling exponent slightly.
This effect is less important in smaller planets that do not achieve high enough pressures.
\\

\section{Super-Mercuries}

The methodology followed for these planets is identical to that for the Earth-like
planets discussed above. We fixed the mineralogy to the best fitting model for the 
Earth (10\% iron end
member and 50\% ferromagnesiowustite in the lower mantle)  and explore
different values for core mass fraction and surface temperature to
understand their effects in the scaling laws. The masses range between
the mass of Mercury and ten mercury-masses.\\

Figure 6 shows the density and temperature structure for a core mass
fraction of 65\% and surface temperature of $440$K corresponding
to Mercury in our solar system (\citealt{Mercury:Internal-Structure}). The purple star
shows data for Mercury.  The discrepancy between one mercury-mass and the planet
mercury comes mostly from the assumptions on the composition (very close to Earth's) and the 
uncertainties in the temperature profile. \\

With these conditions, no Super-Mercury is hot enough in the core
to yield a liquid core and it is clear that high pressure phases in
the mantle appear gradually as the mass of the planet increases. Given our assumptions we 
can not conclude that the planet Mercury in our solar system has a completely
solid core.\\
{[}Figure 6{]}\\

The low central temperatures are a consequence of the relatively small size of the
planet. The pressure at the center of the most massive Super-Mercury is $\sim 260$ GPa. 
Higher surface temperatures would produce a hotter interior.
Figure 7 shows the effect of higher surface temperatures
and different core mass fractions. The exponent for the scaling laws
does not change considerably except for the case in which the surface
temperature is 1500K. Figure 8 shows how in this case, the P-T curves
intersect the solidus yielding about half of the Super-Mercuries with
a liquid and a solid inner core. Thus, the first four planets lie
on a different slope than the next six Super-Mercuries in the core
size and average density figures. The cases for a surface temperature
of 440K and 2000K exhibit a completely solid core or a completely
liquid core respectively. \\
{[}Figure 7, Figure 8{]}\\

The power law exponent to the fits of radius, mantle thickness, core
size and average density with mass for the different cases of surface
temperature and core mass fraction are summarized in Table 2. The
relationship $R\propto M^{\beta}$ between the total radius and mass
 for the Super-Mercuries is closer to the expected $\beta=\frac{1}{3}$
if the average density was constant, than for Super-Earths. We attribute
this effect to the smaller size of the Super-Mercuries that limit
the compressibility and thermal effects on the properties of the
planet.\\
{[}Table 2{]}\\

\section{Conclusions}

The exercise of scaling Earth to larger masses gives an insight into
the internal structure of a massive terrestrial planet. These are
planets that due to their large compressional effects and high internal
temperatures exhibit a mass to radius relationship that deviates from
the cubic power relationship for constant density scaling. For various compositions
the relationship is $R\propto M^{0.267-0.272}$. \\

Super-Mercuries, owing to their small size, do not achieve significant
pressures and temperatures to largely alter the cubic exponent, having
a relationship of $R\propto M^{\sim0.3}$ for different core mass
fractions and surface temperatures. Space missions are biased to detecting
planets with smaller periods, making them closer to its parent star
and hence hotter. It has been shown here that there may be cases where
the surface temperature may have a noticeable effect on the internal density
of the planets. \\

In general, given a pressure regime, the uncertainties in the thermodynamic data
for the mantle and core do not affect the scaling exponents of the bulk properties significantly. 
The same can be said for the uncertainty in the thickness of the bottom boundary layer and core mass fraction.
The pressure regime is mostly controlled by the composition of the planet, therefore different exponents in the scaling
laws would be expected for planets composed of rock and ice/water.\\

It is important to note that the implicit assumption that the Super-Earths
and Super-Mercuries have a thermal evolution similar to Earth,
such that we can use a scaled parameterized convection may restrict
the realm of families for which the scaling laws derived here are
valid. \\

\textbf{ACKNOWLEDGEMENTS}

We would like to thank Sang-Heon Shim for the valuable help on the
topic of equations of state and Lars Stixrude and Carolina Lithgow-Bertollini for
making their thermodynamic data available to us before publication. We are grateful for the useful reviews by 
Christophe Sotin.\\

\bibliographystyle{icarus}
\addcontentsline{toc}{section}{\refname}\bibliography{/home/valencia/ABS/proj}
 \textbf{}\\

\pagebreak 

\textbf{TABLE CAPTIONS}
Table 1.  Thermodynamic data for the core and the mantle used in calculating the density, pressure and temperature profiles of Earth
and the family of SuperEarths.  Data for the mantle is taken from (\citealt{Stixrude_Lithgow:2005}) and different compositions
where tested for the core including extreme end member cases such as FeO and pure Fe for the outer core. Different ratios of Fe and 
$(\text{Mg}_{1-x},\text{Fe}_x)$)  (Wu) are chosen for the mantle with a fixed composition for the core. 
 For this runs, $\gamma$ in the inner 
core has to be chosen with high precision to yield an inner core boundary within 30 Km of the Earth's transition.  The last two entries
correspond to warm geotherms that have an inner core boundary temperature of $\sim 5000$ K that can only be achieved with very large Gruneisen
parameters.  The value of $\gamma_0=2.9$ was chosen for all mantle materials.  
\\

Table 2.  Scaling law exponent of radius, mantle thickness, core radius and average density as a function of mass
for planets with masses ranging from one mercury-mass up to ten times the mass of mercury.  The table shows the effect
of surface temperature and core mass fraction (CMF) in the scaling laws.  Larger CMF yields larger exponents of the scaling
of radius, mantle and core size with mass and a smaller exponent of density with mass.  The effect of the surface temperature
is not monotonic because at $T_surf=440K$ and $T_surf=2000K$ all the planets have either completely solid or liquid cores respectively, but at $T_surf=1500K$ the transition occurs of the first four planets exhibiting a solid core, and the rest a partially liquid core.
\pagebreak

\begin{sidewaystable}
\centering
Table 1\\
\bigskip

\begin{tabular}{|c c|c|c||c|c|c|c|}
\hline 
\multicolumn{2}{|c|}{\textbf{Mantle}} & \textbf{Outer Core} & \textbf{Inner Core} & \multicolumn{4}{|c|} {\textbf{$\beta$}}  \tabularnewline \cline{5-8}
 Fe\% & Wu\% & \footnotesize ($\rho_0, K_{0,300}, {K_{0,300}}', \gamma_0,q, \theta)*$  \normalsize
 & \footnotesize ($\rho_0, K_{T0}, {K_{T0}}', \gamma_0,q, \theta)$ \normalsize  & R & D & R $_{\text{cmb}}$ & $\rho$ \tabularnewline
 \cline{1-2} 
\hline
\hline 
\small 10\% & 30\% \normalsize & FeO         & Fe  & 0.263 & 0.316 & 0.217 & 0.211 \tabularnewline
 & & \footnotesize (6200$^c$,126$^c$,4.8$^c$,2.2$^a$,1.62$^a$,421$^a$) \normalsize  & 
\footnotesize (8171, 135, 6.0, 1.36, 0.91, 998) $^b$ \normalsize & & & & \\ \cline {3-4}
\hline
\small 10\% & 30\% \normalsize  & Fe\%8wSi  & Fe  & 0.272 & 0.293 & 0.252 & 0.186 \tabularnewline
 & & \footnotesize (7793, 141, 5.7, 2.2, 1.62, 421)$^a$ \normalsize  & 
\footnotesize (8171, 135, 6.0, 1.36, 0.91, 998) $^b$ \normalsize & & & & \\ \cline {3-4}
\hline 
 
\small 10\% & 30\% \normalsize & Fe         & Fe  & 0.272 & 0.286 & 0.258 & 0.185 \tabularnewline
 & & \footnotesize (8171, 135, 6.0, 1.36, 0.91, 998) $^b$ \normalsize  & 
\footnotesize (8171, 135, 6.0, 1.36, 0.91, 998) $^b$ \normalsize & & & & \\ \cline {3-4}
\hline 
\hline
\small 10\% & 10\% \normalsize & Fe+alloy    & Fe+alloy  & 0.271 & 0.298 & 0.247 & 0.187 \tabularnewline
 & & \footnotesize (6800$^d$,165$^d$,4.2$^d$,1.9$^e$,1 $^e$,998$^a$) \normalsize & 
\footnotesize (7990$^d$,200$^d$,5.2$^d$,1.734$^d$,1$^e$,998$^a$) \normalsize & & & & \\ \cline {3-4}
\hline 
\small 10\% & 30\% \normalsize & Fe+alloy   & Fe+alloy  & 0.270 & 0.296 & 0.247 & 0.190 \tabularnewline
 & & & \footnotesize  $\gamma = 1.698$ $^d$, $\theta=998$ $^a$ \normalsize & & & & \\ \cline {3-4}
\hline
\small 10\% & 50\% \normalsize & Fe+alloy    & Fe+alloy  & 0.269 & 0.293 & 0.247 & 0.195 \tabularnewline
 & & & \footnotesize  $\gamma = 1.577$, $\theta=464$ $^d$ \normalsize & & & & \\ \cline {3-4}
\hline
\small 0\% & 50\% \normalsize  & Fe+alloy   & Fe+alloy  & 0.269 & 0.293 & 0.247 & 0194 \tabularnewline
 & & & \footnotesize $\gamma = 1.583$ $^d$ \normalsize & & & & \\ \cline {3-4}
\hline
\small 20\% & 50\% \normalsize & Fe+alloy   & Fe+alloy  & 0.268 & 0.294 & 0.247 & 0.195 \tabularnewline
 & & & \footnotesize $\gamma = 1.647$ $^d$, $\theta=998$ $^a$ \normalsize & & & & \\ \cline {3-4}
\hline
\hline
\small 10\% & 30\% \normalsize & Fe+alloy  & Fe+alloy  & 0.271 & 0.300 & 0.244 & 0.188 \tabularnewline
\multicolumn{2}{|c|} {\footnotesize$\gamma_0=2.9, q=1$$^d$} & \footnotesize (6800, 136 ,4.8, 2.5, 1, 421) \normalsize & 
\footnotesize (8051, 217, 4.9, 2.06,0.91, 464)$^d$ \normalsize & & & & \\ \cline {3-4}
\hline
\small 10\% & 30\% \normalsize & Fe+alloy  & Fe+alloy  & 0.271 & 0.299 & 0.247 & 0.186 \tabularnewline
\multicolumn{2}{|c|} {\footnotesize$\gamma_0=2.9, q=1$$^d$} & & 
\footnotesize  (8051, 217, 4.9, 2.338, 1.5, 464)$^d$ \normalsize & & & & \\ \cline {3-4}
\hline
\tabularnewline

\multicolumn{8}{l}{
$^*$ \footnotesize $\rho$ in kg/m$^3$, $K_{0,300}$ in GPa and $\theta$ in Kelvin;} \\
\multicolumn{8}{l}{
\footnotesize $^a$ \citealt{Lin_et_al:2003};} \\
\multicolumn{8}{l}{
\footnotesize $^b$ \citealt{Uchida_et_al:2001};}\\
\multicolumn{8}{l}{
\footnotesize $^c$ \citealt{Williams_Knittle:1997};}\\
\multicolumn{8}{l}{
\footnotesize $^d$ assumed in order to fit density and bulk compressibility to PREM;} \\
\multicolumn{8}{l}{
\footnotesize $^e$ \citealt{Anderson-2000:Melting-Iron}} \\
\normalsize\\

\end{tabular}
\end{sidewaystable}

\pagebreak

Table 2\\

\begin{tabular}{|c|c|c|c|c|c|}
\hline 
T$_{\text{surf}}$&
CMF&
Radius&
Mantle&
Core&
Density\tabularnewline
\hline
\hline 
440&
65&
0.3058&
0.3160&
0.3023&
0.0827\tabularnewline
\hline 
1500&
65&
0.2991&
0.3312&
0.2891&
0.1027\tabularnewline
\hline 
2000&
65&
0.3000&
0.3247&
0.2923&
0.1000\tabularnewline
\hline 
440&
50&
0.3032&
0.3072&
0.3012&
0.0903\tabularnewline
\hline 
440&
80&
0.3094&
0.3371&
0.3042&
0.0718\tabularnewline
\hline
\end{tabular}

\pagebreak

\textbf{FIGURE CAPTIONS}

Figure 1. Schematic temperature profile with conductive gradients
through out the boundary layers at the top and bottom of the mantle
and adiabatic gradients through out the bulk mantle and core. The bottom
boundary layer is chosen to be half the size of the surface boundary
layer.\\

Figure 2. Density (left) and incompressibility (right) of one
earth-mass planets with different compositions. The surface of the
planet is always to the right of the figures. Solid profiles in blue, green
and yellow have 0\%, 10\% and 20\% by mol of Fe end member in the
mantle minerals respectively. The otted line and dahsed line have 0\% and 30\%  
by mol of ferromagnesiowustite in
the lower mantle. The pink and purlple solid lines are the high thermal
profile runs. \\

Figure 3. Density profile (left) for one to ten earth-masses. Each
planet has 32.59\% of its mass in the core and a composition of 10\%
iron in the mantle and 30\% ferromagnesiowustite in the lower mantle.
Power law fit (right) of total radius as a function of mass for ten
different mineral compositions. The red star shows data for Earth.\\

Figure 4. Temperature structure (left) and melting regime (right)
for Super-Earths with a composition of 10\% iron in the mantle and
30\% ferromagnesiowustite in the lower mantle. Solid line shows the geotherm
proposed by (\citealt{Brown_Shankland:1981}). Solidus for a solid
solution of iron and a light element alloy. The Earth is the only
planet that intersects the solidus to yield a liquid outer core. All
other planets have completely solid cores for this particular set of thermodynamic data.\\

Figure 5. Power law fit coefficients of total radius (top left), mantle
thickness (top right), core radius (bottom left) and average density
(bottom right) with mass for Earth-like planets with masses ranging
from 1-10 earth-masses. Blue, green and yellow symbols mean a composition
of 0\%, 10\% and 20\% of Fe in mantle minerals. Open circles, asteriscs
and plus symbols mean 0\%, 30\% and 50\% of ferromangesiowusite with
the rest perovskite in the lower mantle. Pink asterisk represents
a warm geotherm that allows for all Super-Earths to
have a liquid outer core. Purple asterisk has a suitable Lindeman melting law 
that produces completely molten
cores for planets with two to ten earth-masses.\\

Figure 6. Density (right) and temperature (left) profiles of planets
with masses ranging from one to ten mercury-masses with a core mass
fraction of 65\% and 440K surface temperature.\\

Figure 7. Scaling laws for Super-Mercuries for total radius (top left),
mantle thickness (top right), core size (bottom left) and average
density (bottom right). Blue: $\text{T}_{\text{surf}}=440$K \% CMF=65\%,
Green: $\text{T}_{\text{surf}}=440$K \% CMF=50\%, Black: $\text{T}_{\text{surf}}=440$K
\% CMF=80\%, Yellow: $\text{T}_{\text{surf}}=1500$K \% CMF=65\%,
Red: $\text{T}_{\text{surf}}=2000$K \% CMF=65\%.  Purple stars show the
data for the planet Mercury in our solar system. \\

Figure 8. Melting regime for Super-Mercuries with a 65\% core mass
fraction. Three family of planets are shown: with surface temperature
of 2000K that exceeds the solidus, thus all planets have completely
liquid cores; a second family with surface temperature of 1500K that
intersects the solidus yielding 4 planets with liquid cores and 6
with solid inner cores; the last family with a surface temperature
of 440K that exhibits completely frozen cores.\\
\pagebreak

\DeclareGraphicsExtensions{.jpg, .eps}
\DeclareGraphicsRule{.jpg}{eps}{.jpg.bb}{`jpeg2ps -h -r 600 #1}

\textbf{FIGURES}
Figure 1\\
\includegraphics{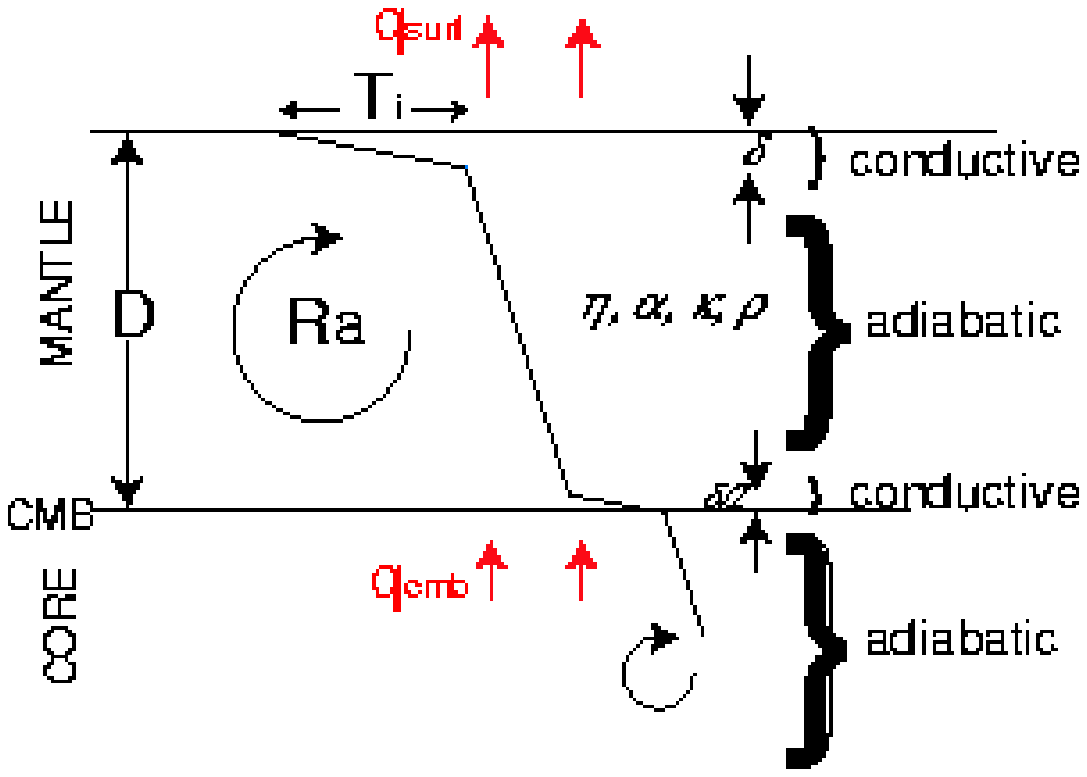}

\pagebreak

Figure 2\\
\includegraphics[%
  scale=0.72,
  angle=90]{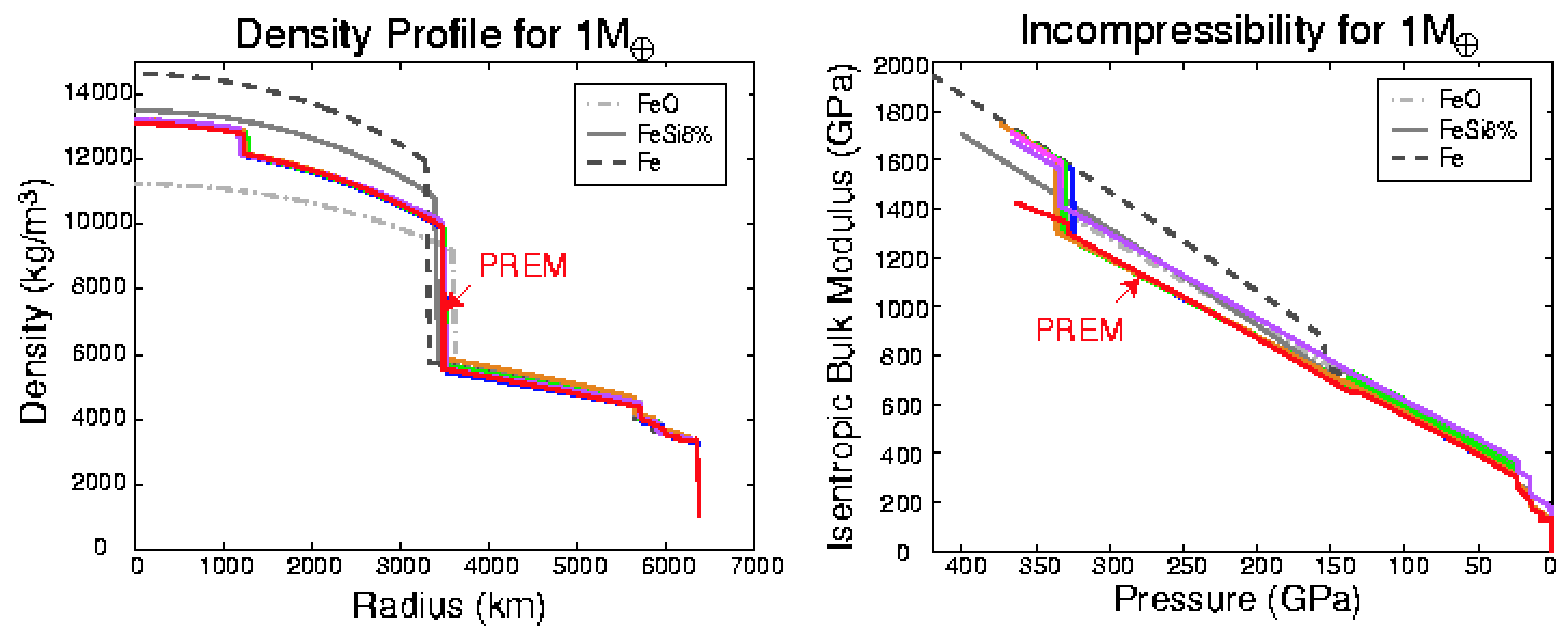}

\pagebreak

Figure 3\\
\includegraphics[%
  scale=0.72,
  angle=90]{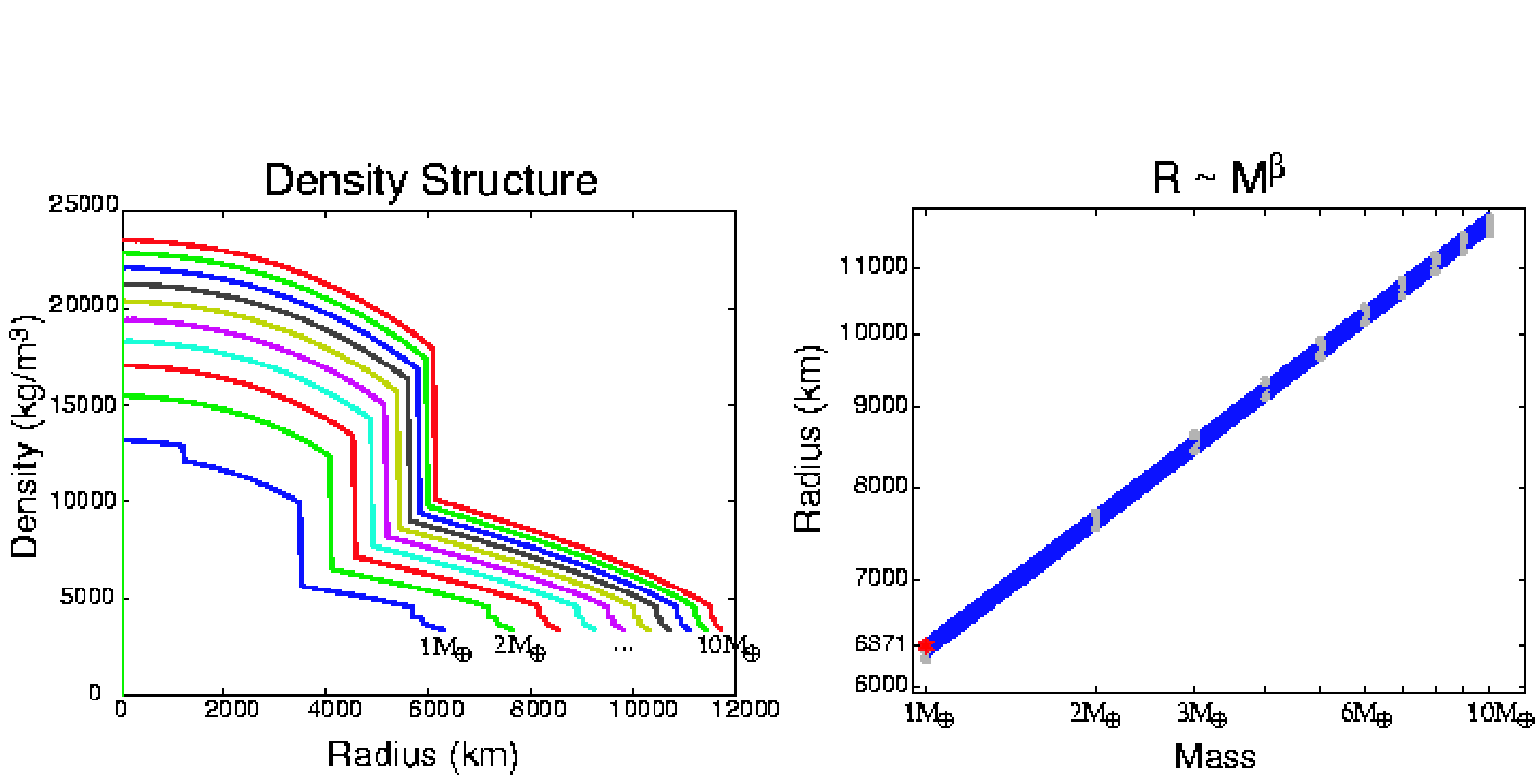}

\pagebreak

Figure 4\\
\includegraphics[%
  scale=0.70,
  angle=90]{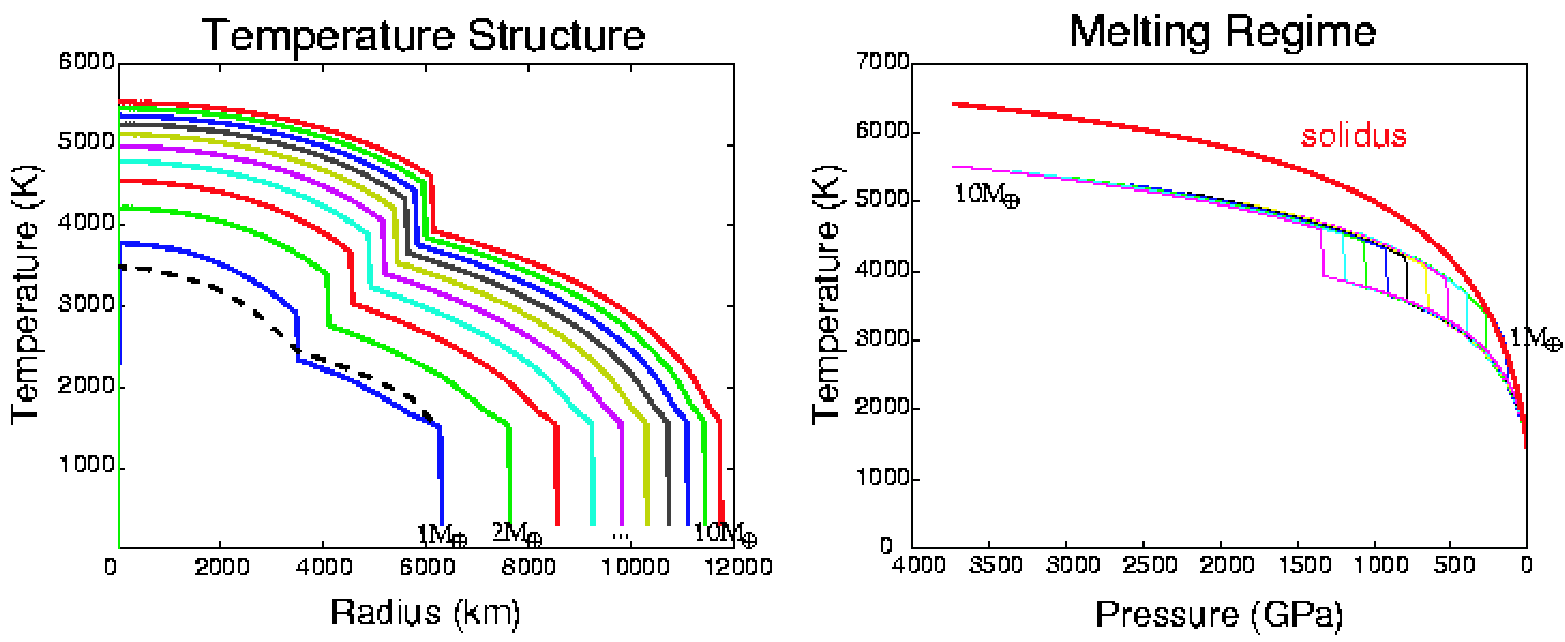}\pagebreak

Figure 5\\
\includegraphics[%
  scale=0.95]{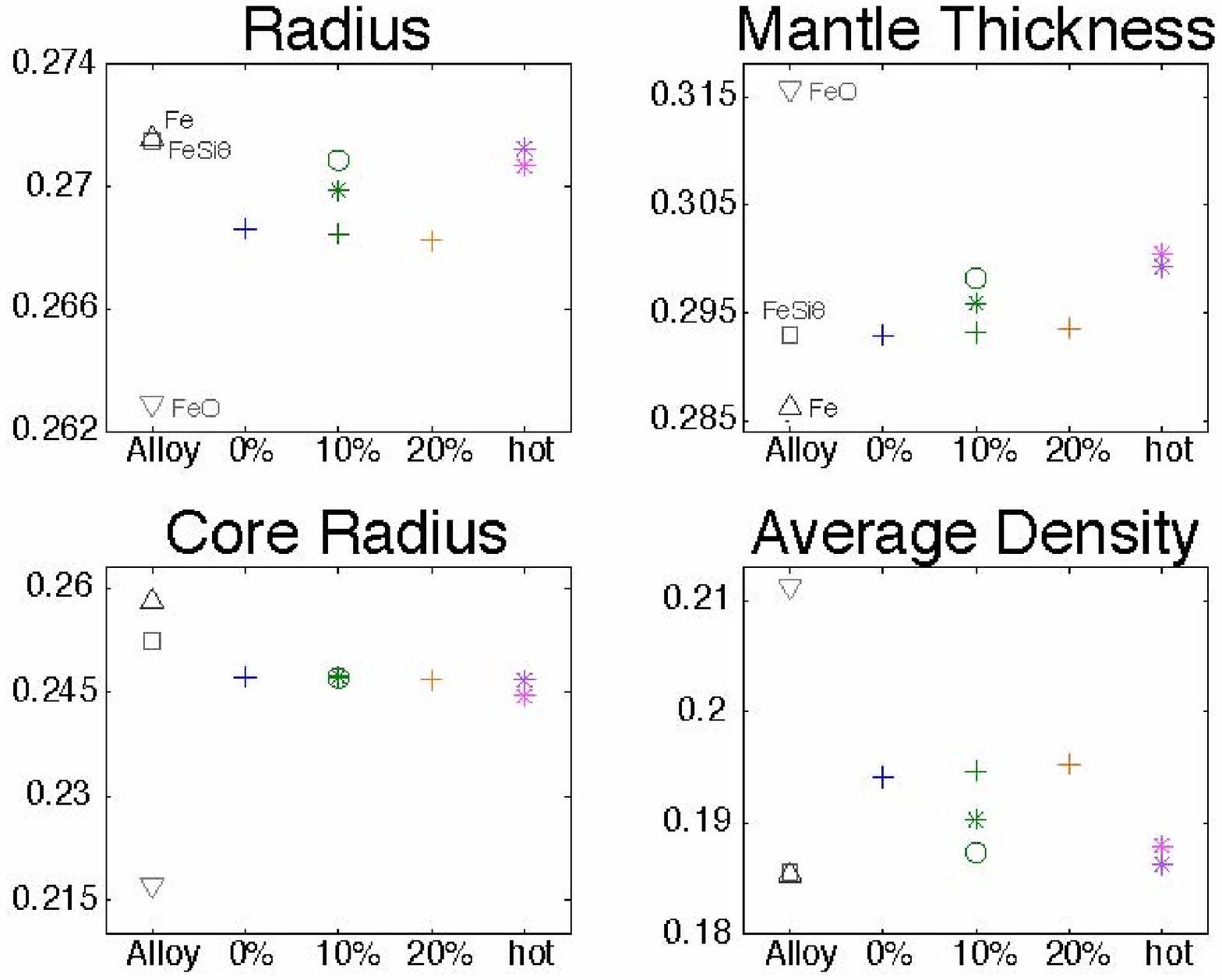}\pagebreak

Figure 6\\
\includegraphics[%
  scale=0.9,
  angle=90]{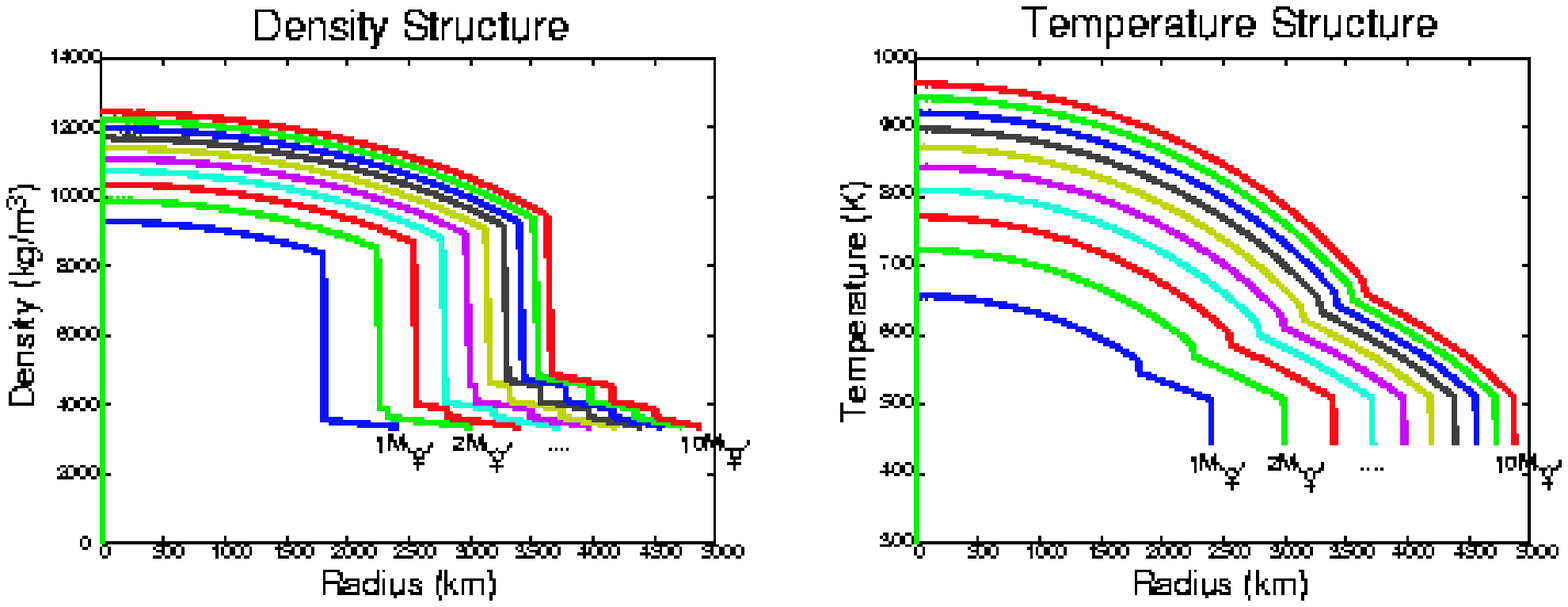}\pagebreak

Figure 7\\
\includegraphics[%
  scale=0.75,
  angle=90]{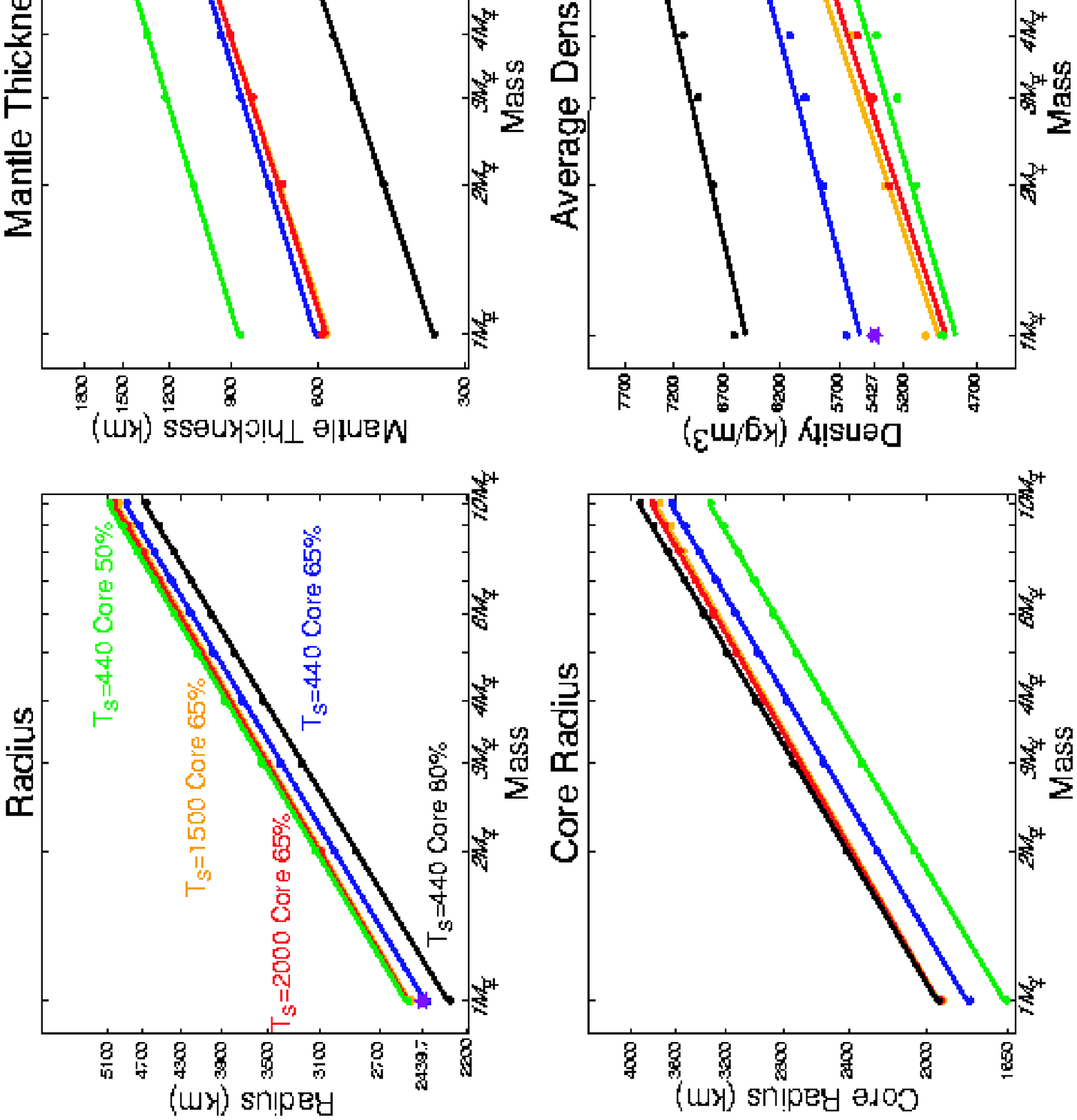}\pagebreak

Figure 8\\
\includegraphics[%
  scale=0.9]{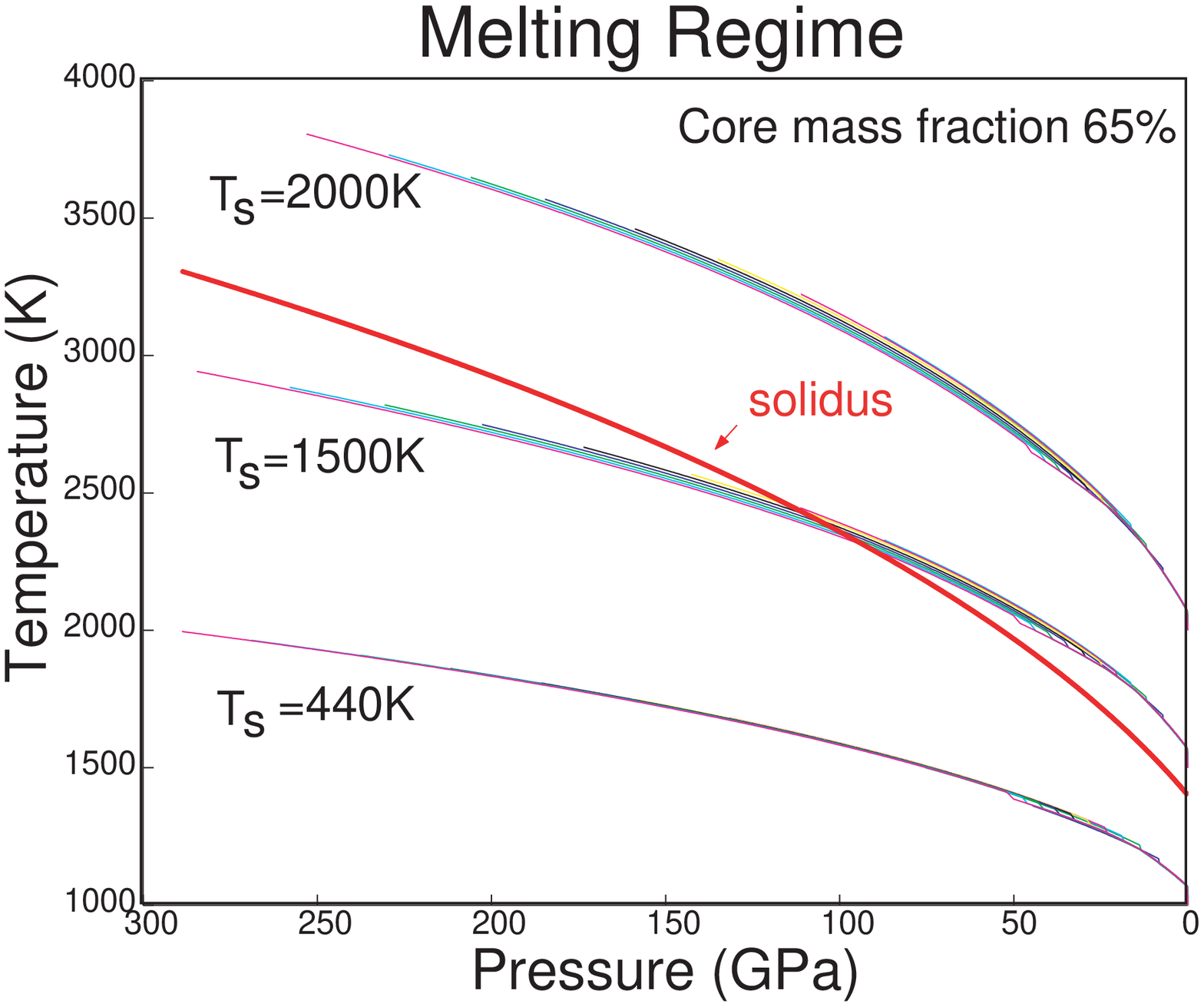}
\end{document}